\documentclass{article}

\PassOptionsToPackage{numbers, compress}{natbib}
\usepackage[final]{neurips_2022_ml4ps}

\usepackage[utf8]{inputenc} 
\usepackage[T1]{fontenc}    
\usepackage{url}            
\usepackage{booktabs}       
\usepackage{amsfonts}       
\usepackage{nicefrac}       
\usepackage{xcolor}         
\usepackage{graphicx}
\usepackage{amsmath}

\usepackage{siunitx}
\DeclareSIUnit\parsec{pc}
\DeclareSIUnit\years{yr}
\DeclareSIUnit\Msol{M_{\odot}}
\DeclareSIUnit\Lsol{L_{\odot}}
\DeclareSIUnit\AU{au}
\DeclareSIUnit\om{\Omega}
\DeclareSIUnit\orb{T_{\mathrm{orb}}}
\DeclareSIUnit\scaleheight{H}
\newcommand{\coala}{\texttt{COALA}}

\title{A Neural Network Subgrid Model of the Early Stages of Planet Formation}

\author{Thomas Pfeil$^{*, 1,2}$ \,\, Miles Cranmer$^{3,4}$  \,\, Shirley Ho$^{3,4,5,6}$ \,\, Philip J. Armitage$^{4,7}$  \\ \bfseries Tilman Birnstiel$^{1,8}$  \,\, Hubert Klahr$^2$ \\ \\
$^*$ \href{mailto:tpfeil@usm.lmu.de}{tpfeil@usm.lmu.de}\\
$^1$ University Observatory, Ludwig-Maximilians-Universität München, Munich, Germany \\
$^2$ Max Planck Institute for Astronomy, Heidelberg, Germany \\
$^3$ Department for Astrophysical Sciences, Princeton University, Princeton, USA \\
$^4$ Center for Computational Astrophysics, Flatiron Institute, New York, USA \\
$^5$ Center for Cosmology and Particle Physics, New York University, New York, USA\\
$^6$ Department of Physics, Carnegie Mellon University, Pittsburgh, USA \\
$^7$ Department of Physics and Astronomy, Stony Brook University, Stony Brook, USA \\
$^8$ Exzellenzcluster ORIGINS, Boltzmannstr. 2, D-85748 Garching, Germany
}

\usepackage{hyperref} 
\begin{document}

\maketitle

\begin{abstract}
Planet formation is a multi-scale process in which the coagulation of \si{\micro \meter}-sized dust grains in protoplanetary disks is strongly influenced by the hydrodynamic processes on scales of astronomical units ($\approx\SI{1.5e8}{\kilo \meter}$).
Studies are therefore dependent on subgrid models to emulate the micro physics of dust coagulation on top of a large scale hydrodynamic simulation.
Numerical simulations which include the relevant physical effects are complex and computationally expensive. 
Here, we present a fast and accurate learned effective model for dust coagulation, trained on data from high resolution numerical coagulation simulations. Our model captures details of the dust coagulation process that were so far not tractable with other dust coagulation prescriptions with similar computational efficiency.
\end{abstract}

\section{Introduction to Dust Coagulation - The First Stage of Planet Formation}
After the formation of a protostar, remaining material of its parent molecular cloud core forms a so-called protoplanetary disk around it. About \SI{1}{\percent} of the mass of this disk consists of solids in the form of initially \si{\micro\meter}-sized carbonaceous silicate grains and ices. All solid objects, including the rocky planets, the rocky cores of gas giant planets, comets, and asteroids form out of this material. Subsequent collisions between the grains are caused by gas turbulence and differential aerodynamic drag and lead to the formation of larger aggregates via sticking due to van der Waals forces. Since relative velocities between the grains increase with their sizes, growth is halted at some point, when collisions become too violent for sticking and instead lead to fragmentation (break-up). At this so-called fragmentation barrier, an equilibrium size distribution is reached. Its form is determined by the interior composition of the grains and their size-dependent relative velocities. 

Theoretically, these processes are described by the Smoluchowski equation \cite{Smoluchowski1916}---an integro-differential equation that gives the mass exchange rates between grains on a continuous spectrum of sizes. Only a few analytically solvable cases exist, which is why most numerical models of dust coagulation rely on solution techniques for the discretized Smoluchowski equation, which is derived by exchanging the continuum of grain sizes by a discreet grid of sizes. Solving the resulting system of ODEs is an elaborate numerically task that requires the size grid to have >100 bins to lead to meaningful results \citep{Brauer2008, Birnstiel2009}. 

\begin{figure}
    \centering
    \includegraphics[width=\textwidth]{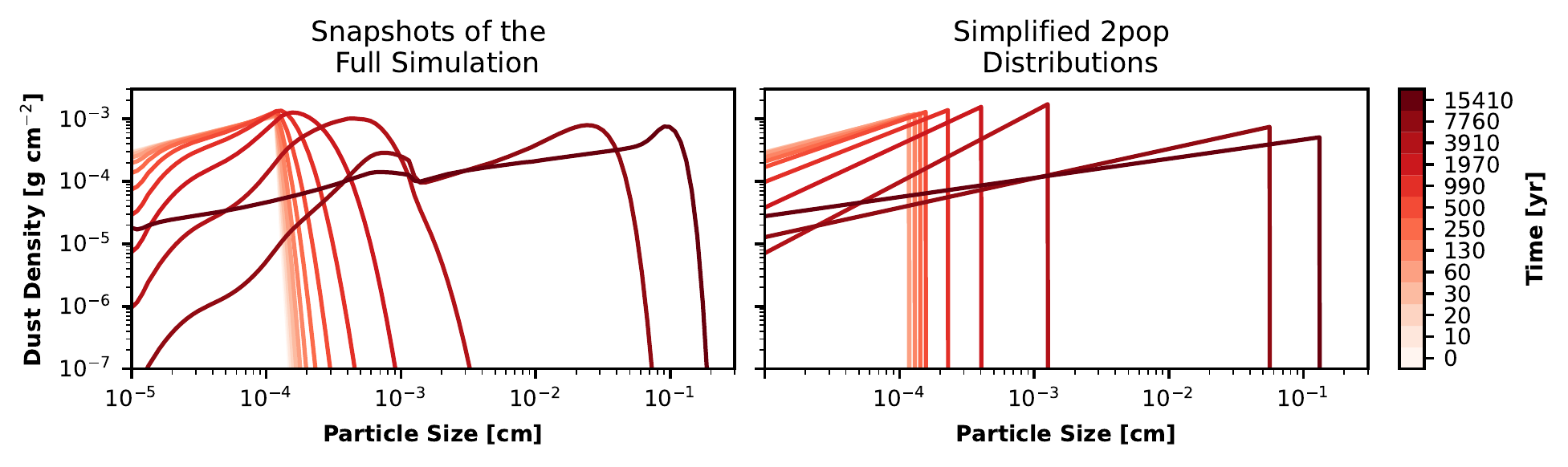}
    \vspace{-0.3cm}
    \caption{Output of a numerical simulation of dust coagulation in a protoplanetary disk (left side). Initially \si{\micro \meter}-sized grains grow until they reach the fragmentation barrier. On the right-hand side we show the equivalent power law size distributions derived from the actual simulation results on the left. The simplified time series data is the training data for our machine learning model.}
    \label{fig:2popcompare}
\end{figure}

An example simulation is shown in the left hand side of \autoref{fig:2popcompare}. The model is initialized with a distribution of \si{\micro \meter}-sized grains. Collisions first lead to an almost exponential growth phase, which, in this case is halted by fragmentation after $\sim \SI{{}e4}{\years}$ . The result is a top-heavy equilibrium distribution of up to \SI{2}{\milli \meter}-sized grains.

These multi-bin models are applicable to 0D \citep[local; see][]{Brauer2008} or 1D \citep[vertically and azimuthally averaged; see][]{Stammler2022} disk models, but due to their high numerical cost, can not be applied in 3D models of protoplanetary disks.  

\subsection{A Power Law Prescription for Dust Coagulation and the Need for a Machine Learning Approach}

We aim to develop an approach in which the dust size distribution is described by a truncated power law, instead of a discretized distribution with hundreds of size bins. Our goal is to make the modeling of dust coagulation on top of large scale hydrodynamic simulations more feasible.
For a given total dust column density $\sigma_\text{tot}$, and a minimum particle size $a_\text{min}=\SI{{}e-5}{\cm}$, this simplified distribution can be described by only two parameters:
\begin{align*}
    a_\text{max}: \quad &\text{The size of the largest particles (truncation size of the power law)} \\ 
    \sigma_1: \quad &\text{The column density of particles larger than $a_\text{int}=\sqrt{a_\text{max}a_\text{min}}$}.
\end{align*}
It can be shown that the exponent of the power law size distribution $\sigma(a)\propto a^{p+4}$ is then given by $
    p = \frac{\log{\left(\sigma_1/\sigma_0\right)}}{\log{\left(a_\text{max}/a_\text{int}\right)}} - 4 $,
where $\sigma_0= \sigma_\text{tot}-\sigma_1$ is the column density of particles smaller than $a_\text{int}$.
In contrast to other approximate models like \texttt{two-pop-py} \citep{Birnstiel2012}, this approach makes it possible to retain information about the overall shape of the size distribution.
It is, however, not trivial to find a mathematical description for the time evolution of the power law distribution  without making strongly simplifying assumptions. 
We therefore propose a machine learning aided power law model, which predicts the time evolution of the simplified distribution.

\section{Method}
\begin{figure}
    \centering
    \includegraphics[width=\textwidth]{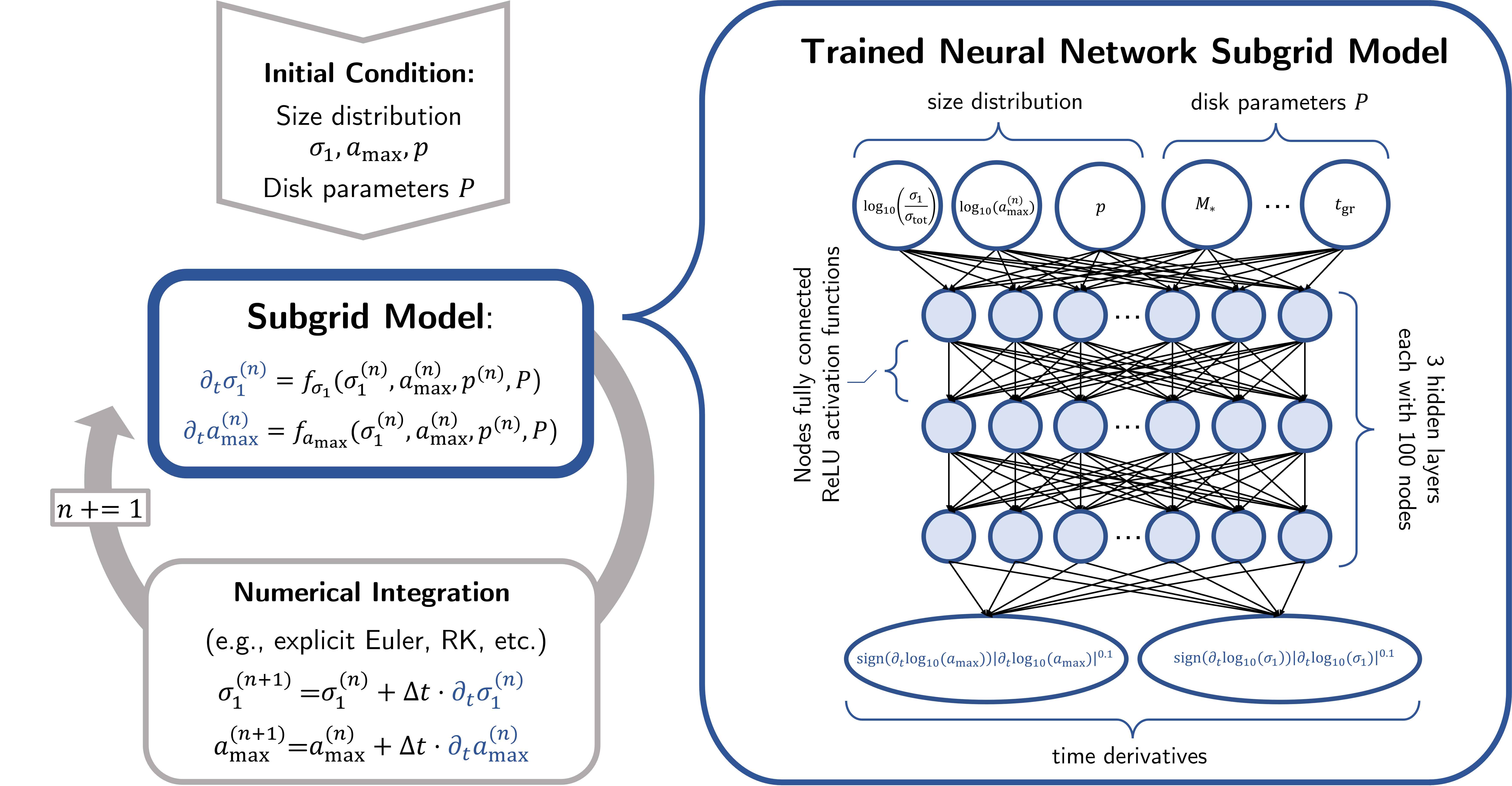}
    \caption{General outline of the trained machine learning subgrid model of dust coagulation. An artificial neural network is trained to predict the time derivatives of the size distribution's power law representation. The resulting source term is used to evolve the distribution in time.}
    \label{fig:method}
\end{figure}

For our method, we trained a Multilayer Perceptron (MLP) on the evolution of power law grain size distributions derived from detailed multi-bin simulations of dust coagulation. The general workflow of our model is laid out in  \autoref{fig:method}. 
The inputs of our neural network are the size distribution parameters, and the parameters of the protoplanetary disk environment, like gas temperature, gas density, etc. The model's output are the respective time derivatives $\partial_t a_\text{max}$ and $\partial_t \sigma_1$, which are then used as source terms in a numerical integration scheme. 
\\
Our simple neural network model therefore makes it possible to simulate the temporal evolution of the physical system, similar to other machine learning approaches explored in recent years \cite{Sanchez-Gonzales2020, Kidger2022}.
\\
Our MLP consists of 3 hidden layers, each with 100 nodes, 14 nodes in the input layer, and two nodes in the output layer. The layers are fully connected with ReLU activation functions.

\subsection{Training Data Generation}
We create our training data using the \coala{} dust coagulation routine, which was provided by Til Birnstiel and Sebastian Stammler, and which was already used in a hydrodynamic simulation \cite{Drazkowska2019}. 
\coala{} is a local dust coagulation code, written in FORTRAN that numerically solves the Smoluchowski equation on a mass grid (in our case with 171 bins). 10000 dust coagulation simulations have been created, each with 150 time outputs. The dust distributions are evolved over a time corresponding to 50 dust growth time scales, or maximally \SI{{}e6}{\years} to ensure that an equilibrium is reached at the end of each simulation.
The initial conditions are chosen randomly from a parameter space that represents the known typical conditions within protoplanetary disks from simulations and observations.

\subsection{Training Data Pre-Processing}
As a first step, we derive the two parameters of the power law size distributions from the full size distributions with 171 size bins. We define $a_\text{max}$ as the particle size for which $\int_{a_\text{min}}^{a_\text{max}} \sigma(a) \, \text{d}a /\sigma_\text{tot} = 0.99$ holds, i.e. \SI{99}{\percent} of the total mass of the particles has sizes smaller than $a_\text{max}$. $\sigma_1$ is then derived by summing up the mass of all bins with sizes larger than $a_\text{int}=\sqrt{a_\text{max}a_\text{min}}$.
This results in 10000$\times$150 time series data points for both quantities, from which we derive the respective time derivatives. 
For training, we scale the data to a range from 0 to 1 and divide the dataset into 8000 training data simulations and 2000 test data simulation. We found that even small deviations from the actual equilibrium states can lead to large errors after time integration with the predicted gradients. Our experiments have shown that the best training results are achieved if we use the tenth root of the time derivatives, multiplied by their sign as the training data. In that way, also small scale features around the equilibrium states ($\partial_t=0$) can be learned, leading to the best results during numerical integration and to the correct equilibrium distribution.

\subsection{Training Procedure}
We train our neural network model within the Pytorch Lightning framework \cite{Pasze2019, Falcon2020}, using the Adam optimization algorithm \cite{Kingma2017}. The batch size is set to 1000, we apply a learning rate of $3\times 10^{-4}$, and train the model for 1000 epochs. We employ the Mean Absolute Percentage Error  \cite[MAPE,][]{Myttenaere2016} as a loss function, which also penalized deviations of small absolute value. To avoid division by zero when applying the loss function, we offset the normalized training data by +0.1. Training was conducted on a single Nvidia A100-40GB GPU.

\section{Results}\label{sec:results}
\begin{figure}[ht]
    \centering
    \includegraphics[width=\textwidth]{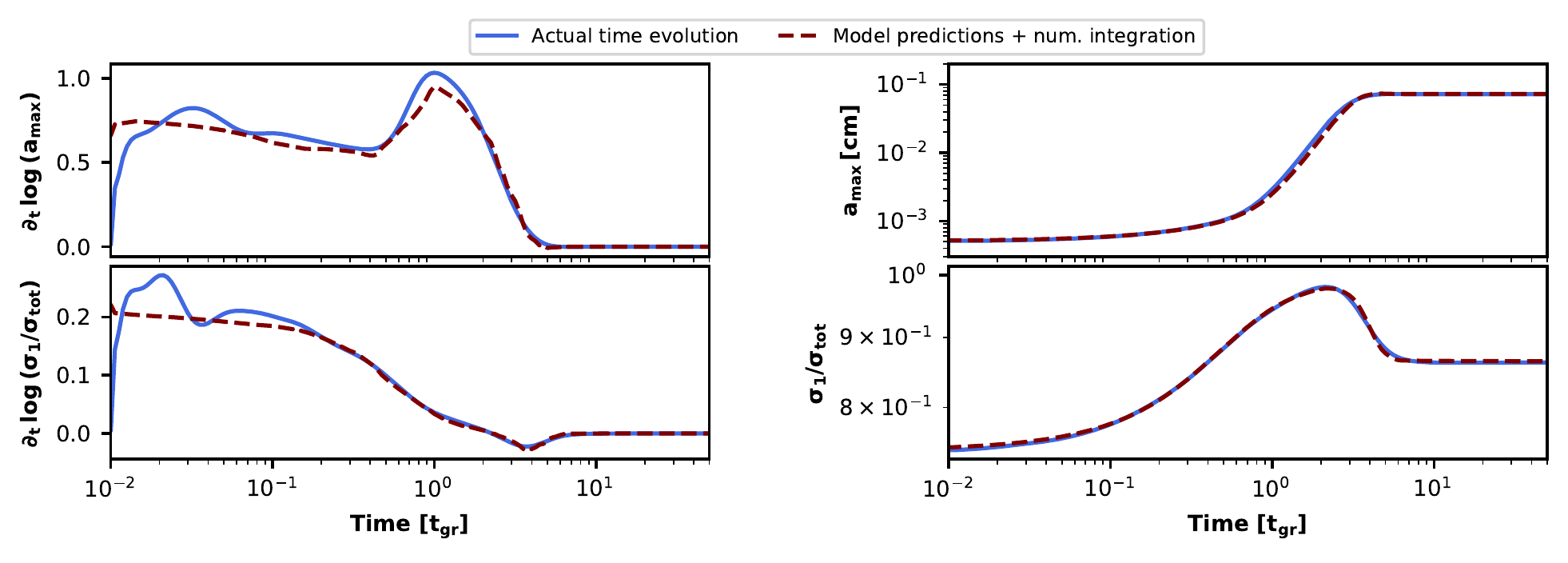}
    \vspace{-0.5cm}
    \caption{Result of a numerical integration with the neural network predictions for the respective time derivatives.}
    \label{fig:result}
\end{figure}

After training we evaluate the resulting model by using the predicted time derivatives for numerical time integration of the setups from the training data set (see our method in \autoref{fig:method} and  \url{https://github.com/ThomasPfeil/2popML}). 
For the tests performed in this work, we utilize an explicit Euler scheme, as shown in \autoref{fig:method}. We limit the time step to ensure numerical stability for the given source terms as
\begin{equation}
    \Delta t = C\cdot\min\left(\left|\frac{a_\text{max}}{\partial_t a_\text{max}}\right|, \left|\frac{\sigma_1}{\partial_t\sigma_1}\right|\right),
\end{equation}
with $C=0.1$. In \autoref{fig:result}, we present an example simulation from the test dataset. 
The average deviation from the actual time series is about $\sim \SI{4}{\percent}$. We have conducted this procedure with all 2000 parameter combinations from the test data set. 
On average, one full integration run takes $\approx\SI{73}{\milli \second}$ wall clock time, compared to \SI{791}{\milli \second} for the full numerical model on the same machine. 
11 integrations failed, reaching either negative dust densities or errors larger than \SI{1000}{\percent}, resulting in a \SI{99.45}{\percent} success rate.

\begin{figure}[ht]
    \centering
    \includegraphics[width=\textwidth]{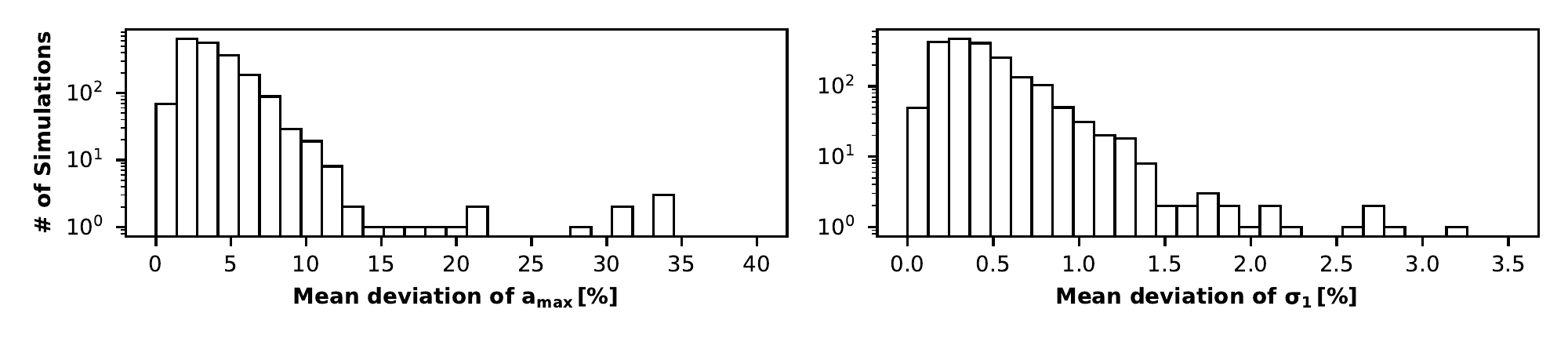}
    \vspace{-0.5cm}
    \caption{Distribution of deviations from the actual simulation time series for both simulated parameters $a_\text{max}$ and $\sigma_1$.}
    \label{fig:error}
\end{figure}

For the 1989 successfully finished test simulations, we plot the distribution of the mean relative deviation of each time series to the respective actual time series in \autoref{fig:error}. On average, the deviation between the integration series conducted with the model prediction and the actual data is $\sim \SI{4}{\percent}$ for the maximum particle size, and $\sim\SI{0.5}{\percent}$ for the column density of large particles.

\vspace{-0.3cm}
\section{Conclusions and Outlook}
Our results strongly suggest that numerical efforts to study the early phases of planet formation can benefit from the use of machine learning techniques. 
Our neural network model was capable of predicting gradients with high enough precision to allow for time integration of the vast majority of the test data set (\SI{99.45}{\percent} of the simulations).
Our model could therefore be used as a fast and accurate alternative to commonly used full coagulation simulations. Due to its much shorter runtime, it could, for the first time, make large scale hydrodynamic simulations of protoplanetary disks with dust coagulation feasible. 

Since our model is trained on simulation data with various parameter combinations, we expect it to produce accurate results as long as the applied model parameters lie within the ranges used for training. This means, the most important limitation of our model lies in the range of applicable stellar parameters and disk parameters, e.g., stellar mass (varies from 0.01 to \SI{1.4}{\Msol}), distances to the central star (varied from 0.1 to \SI{100}{\AU}), etc.

Further testing is needed for the use of our model in disks with substructure, e.g., disks with planetary gaps and pressure bumps. It is not clear if our model will produce reliable outputs in these environments, since it was trained on parameter combinations derived from simple power law disks (without substructure).

Testing this requires an implementation of our neural network model as a subgrid model into a hydrodynamics code to simulate gas and dust dynamics in protoplanetary disks. We therefore aim to couple our model to the \texttt{PLUTO} code \citep{Mignone2007}.
Once a stable run is achieved, we can test our subgrid model in an evolving environment and under the conditions in substructures.

The structural similarity of our approach (\autoref{fig:method}) to semi-analytic physical models could also make it possible to interpret the trained neural network in the future and derive insights into the underlying physics, which could make our results interpretable \cite{Cranmer2020, Kochov2021, Stachenfeld2021}.

\section*{Acknowledgments}
T.P. expresses his gratitude to the Simons Foundation for the opportunity to conduct this project as part of the 2022 Flatiron Machine Learning X Science Summer School. Special thanks goes to the summer school mentors S.H., M.C., and P.A. for their advise and many helpful discussions. 
T.P., H.K., and T.B. acknowledge the support of the German Science Foundation (DFG) priority program SPP 1992 “Exploring the Diversity of Extrasolar Planets” under grant Nos. BI 1816/7-2 and KL 1469/16-1/2. 
T.B. acknowledges funding from the European Research Council (ERC) under the European Union’s Horizon 2020 research and innovation programme under grant agreement No 714769 and funding by the Deutsche Forschungsgemeinschaft (DFG, German Research Foundation) under grants 361140270, 325594231, and Germany’s Excellence Strategy - EXC-2094 - 390783311.
All computations were conducted on the VERA cluster of the Max Planck Institute for Astronomy, Heidelberg.

\bibliography{Literature}

\begin{thebibliography}{16}
\providecommand{\natexlab}[1]{#1}
\providecommand{\url}[1]{\texttt{#1}}
\expandafter\ifx\csname urlstyle\endcsname\relax
  \providecommand{\doi}[1]{doi: #1}\else
  \providecommand{\doi}{doi: \begingroup \urlstyle{rm}\Url}\fi

\bibitem[{Smoluchowski}(1916)]{Smoluchowski1916}
M.~V. {Smoluchowski}.
\newblock {Drei Vortrage uber Diffusion, Brownsche Bewegung und Koagulation von
  Kolloidteilchen}.
\newblock \emph{Zeitschrift fur Physik}, 17:\penalty0 557--585, January 1916.

\bibitem[{Brauer} et~al.(2008){Brauer}, {Dullemond}, and {Henning}]{Brauer2008}
F.~{Brauer}, C.~P. {Dullemond}, and Th. {Henning}.
\newblock {Coagulation, fragmentation and radial motion of solid particles in
  protoplanetary disks}.
\newblock \emph{A\&A}, 480\penalty0 (3):\penalty0 859--877, March 2008.
\newblock \doi{10.1051/0004-6361:20077759}.

\bibitem[{Birnstiel} et~al.(2009){Birnstiel}, {Dullemond}, and
  {Brauer}]{Birnstiel2009}
T.~{Birnstiel}, C.~P. {Dullemond}, and F.~{Brauer}.
\newblock {Dust retention in protoplanetary disks}.
\newblock \emph{A\&A}, 503\penalty0 (1):\penalty0 L5--L8, August 2009.
\newblock \doi{10.1051/0004-6361/200912452}.

\bibitem[{Stammler} and {Birnstiel}(2022)]{Stammler2022}
Sebastian~M. {Stammler} and Tilman {Birnstiel}.
\newblock {DustPy: A Python Package for Dust Evolution in Protoplanetary
  Disks}.
\newblock \emph{ApJ}, 935\penalty0 (1):\penalty0 35, August 2022.
\newblock \doi{10.3847/1538-4357/ac7d58}.

\bibitem[{Birnstiel} et~al.(2012){Birnstiel}, {Klahr}, and
  {Ercolano}]{Birnstiel2012}
T.~{Birnstiel}, H.~{Klahr}, and B.~{Ercolano}.
\newblock {A simple model for the evolution of the dust population in
  protoplanetary disks}.
\newblock \emph{A\&A}, 539:\penalty0 A148, March 2012.
\newblock \doi{10.1051/0004-6361/201118136}.

\bibitem[{Sanchez-Gonzalez} et~al.(2020){Sanchez-Gonzalez}, {Godwin}, {Pfaff},
  {Ying}, {Leskovec}, and {Battaglia}]{Sanchez-Gonzales2020}
Alvaro {Sanchez-Gonzalez}, Jonathan {Godwin}, Tobias {Pfaff}, Rex {Ying}, Jure
  {Leskovec}, and Peter~W. {Battaglia}.
\newblock {Learning to Simulate Complex Physics with Graph Networks}.
\newblock \emph{arXiv e-prints}, art. arXiv:2002.09405, February 2020.

\bibitem[{Kidger}(2022)]{Kidger2022}
Patrick {Kidger}.
\newblock {On Neural Differential Equations}.
\newblock \emph{arXiv e-prints}, art. arXiv:2202.02435, February 2022.

\bibitem[{Dr{\k{a}}{\.z}kowska} et~al.(2019){Dr{\k{a}}{\.z}kowska}, {Li},
  {Birnstiel}, {Stammler}, and {Li}]{Drazkowska2019}
Joanna {Dr{\k{a}}{\.z}kowska}, Shengtai {Li}, Til {Birnstiel}, Sebastian~M.
  {Stammler}, and Hui {Li}.
\newblock {Including Dust Coagulation in Hydrodynamic Models of Protoplanetary
  Disks: Dust Evolution in the Vicinity of a Jupiter-mass Planet}.
\newblock \emph{ApJ}, 885\penalty0 (1):\penalty0 91, November 2019.
\newblock \doi{10.3847/1538-4357/ab46b7}.

\bibitem[{Paszke} et~al.(2019){Paszke}, {Gross}, {Massa}, {Lerer}, {Bradbury},
  {Chanan}, {Killeen}, {Lin}, {Gimelshein}, {Antiga}, {Desmaison}, {K{\"o}pf},
  {Yang}, {DeVito}, {Raison}, {Tejani}, {Chilamkurthy}, {Steiner}, {Fang},
  {Bai}, and {Chintala}]{Pasze2019}
Adam {Paszke}, Sam {Gross}, Francisco {Massa}, Adam {Lerer}, James {Bradbury},
  Gregory {Chanan}, Trevor {Killeen}, Zeming {Lin}, Natalia {Gimelshein}, Luca
  {Antiga}, Alban {Desmaison}, Andreas {K{\"o}pf}, Edward {Yang}, Zach
  {DeVito}, Martin {Raison}, Alykhan {Tejani}, Sasank {Chilamkurthy}, Benoit
  {Steiner}, Lu~{Fang}, Junjie {Bai}, and Soumith {Chintala}.
\newblock {PyTorch: An Imperative Style, High-Performance Deep Learning
  Library}.
\newblock \emph{arXiv e-prints}, art. arXiv:1912.01703, December 2019.

\bibitem[{Falcon}(2019)]{Falcon2020}
W.A. {Falcon}.
\newblock {PyTorch Lightning}.
\newblock \emph{Github, \url{https://github.com/Lightning-AI/lightning}}, March
  2019.
\newblock \doi{10.5281/zenodo.3828935}.

\bibitem[{Kingma} and {Ba}(2014)]{Kingma2017}
Diederik~P. {Kingma} and Jimmy {Ba}.
\newblock {Adam: A Method for Stochastic Optimization}.
\newblock \emph{arXiv e-prints}, art. arXiv:1412.6980, December 2014.

\bibitem[{De Myttenaere} et~al.(2016){De Myttenaere}, {Golden}, {Le Grand}, and
  {Rossi}]{Myttenaere2016}
Arnaud {De Myttenaere}, Boris {Golden}, B{\'e}n{\'e}dicte {Le Grand}, and
  Fabrice {Rossi}.
\newblock {Mean Absolute Percentage Error for regression models}.
\newblock \emph{arXiv e-prints}, art. arXiv:1605.02541, May 2016.

\bibitem[{Mignone} et~al.(2007){Mignone}, {Bodo}, {Massaglia}, {Matsakos},
  {Tesileanu}, {Zanni}, and {Ferrari}]{Mignone2007}
A.~{Mignone}, G.~{Bodo}, S.~{Massaglia}, T.~{Matsakos}, O.~{Tesileanu},
  C.~{Zanni}, and A.~{Ferrari}.
\newblock {PLUTO: A Numerical Code for Computational Astrophysics}.
\newblock \emph{ApJs}, 170:\penalty0 228--242, May 2007.
\newblock \doi{10.1086/513316}.

\bibitem[{Cranmer} et~al.(2020){Cranmer}, {Sanchez-Gonzalez}, {Battaglia},
  {Xu}, {Cranmer}, {Spergel}, and {Ho}]{Cranmer2020}
Miles {Cranmer}, Alvaro {Sanchez-Gonzalez}, Peter {Battaglia}, Rui {Xu}, Kyle
  {Cranmer}, David {Spergel}, and Shirley {Ho}.
\newblock {Discovering Symbolic Models from Deep Learning with Inductive
  Biases}.
\newblock \emph{arXiv e-prints}, art. arXiv:2006.11287, June 2020.

\bibitem[Kochkov et~al.(2021)Kochkov, Smith, Alieva, Wang, Brenner, and
  Hoyer]{Kochov2021}
Dmitrii Kochkov, Jamie~A. Smith, Ayya Alieva, Qing Wang, Michael~P. Brenner,
  and Stephan Hoyer.
\newblock Machine learning\&\#x2013;accelerated computational fluid dynamics.
\newblock \emph{Proceedings of the National Academy of Sciences}, 118\penalty0
  (21):\penalty0 e2101784118, 2021.
\newblock \doi{10.1073/pnas.2101784118}.
\newblock URL \url{https://www.pnas.org/doi/abs/10.1073/pnas.2101784118}.

\bibitem[{Stachenfeld} et~al.(2021){Stachenfeld}, {Fielding}, {Kochkov},
  {Cranmer}, {Pfaff}, {Godwin}, {Cui}, {Ho}, {Battaglia}, and
  {Sanchez-Gonzalez}]{Stachenfeld2021}
Kimberly {Stachenfeld}, Drummond~B. {Fielding}, Dmitrii {Kochkov}, Miles
  {Cranmer}, Tobias {Pfaff}, Jonathan {Godwin}, Can {Cui}, Shirley {Ho}, Peter
  {Battaglia}, and Alvaro {Sanchez-Gonzalez}.
\newblock {Learned Coarse Models for Efficient Turbulence Simulation}.
\newblock \emph{arXiv e-prints}, art. arXiv:2112.15275, December 2021.

\end{thebibliography}

\end{document}